\documentclass[]{article}
\usepackage[margin=1.5in]{geometry}
\usepackage[english]{babel}
\usepackage[utf8x]{inputenc}
\usepackage{amsmath}
\usepackage{graphicx}
\usepackage{setspace}
\usepackage{lineno}
\usepackage{cite}

\title{Coefficient-of-Determination Fourier Transform }
\author{Matthew Marko}
\date{1 August 2018}

\begin{document}

\maketitle

\abstract{This algorithm is designed to perform numerical transforms to convert data from the temporal domain into the spectral domain.  This algorithm obtains the spectral magnitude and phase by studying the Coefficient of Determination of a series of artificial sinusoidal functions with the temporal data, and normalizing the variance data into a high-resolution spectral representation of the time-domain data with a finite sampling rate.  What is especially beneficial about this algorithm is that it can produce spectral data at any user-defined resolution, and this highly resolved spectral data can be transformed back to the temporal domain.  }

\noindent \emph{Naval Air Warfare Center Aircraft Division}, \\
Joint-Base McGuire-Dix-Lakehurst, Lakehurst NJ 08733, USA.  NAVAIR Public Release 2016-755  Distribution Statement A - "Approved for public release; distribution is unlimited."

\section{{Introduction}}

The \emph{Fourier Transform} \cite{ODEbook,PDEbook,01455106, josab-17-10-1795, MathMethodsBook, EngAnalysisBook,FFTbook} is one of the most widely used mathematical operators in all of engineering and science \cite{01093941,a15622,a27715}.  The Fourier Transform can take a temporal function and convert it into a series of sinusoidal functions, offering significant clarity on the nature of the data.  While the original Fourier Transform is an analytical mathematical operator, \emph{Discrete Fourier Transform} (DFT) methods are overwhelmingly used to take incoherent temporal measurements and convert them into spectral plots based on real, experimental data.  

The author proposes a numerical algorithm to perform a highly-resolved spectral transform of a temporal function of limited resolution.  The spectral magnitude is determined by finding the magnitude of the Coefficient of Determination of the function as compared with a given sinusoidal function; this represents the independent spectral value as a function of the sinusoidal frequency.  Rather than the spectral domain being proportional to the time step, the user defines exactly which frequencies are necessary to investigate.  The spectral domain can be as large or as resolved as is necessary; the resolution possible is limited only by the abilities of the computer performing the transform.  

\section{{Spectral Transform}}

The transform starts by first determining the peak total range of the data in the temporal domain, this range will become the base amplitude of the spectral series.  The computer then generates a series of sine and cosine functions at each frequency within the spectral domain, and compares each of these sinusoidal functions to the temporal data to be transformed.  In the comparison, a correlation coefficient is found and saved.  To accommodate fluctuations in phase, each frequency generates both a sine and cosine function; this ultimately results in real and imaginary spectral components.  Finally, the magnitudes of the correlation factor data is normalized, and the result is an accurate spectral representation of the temporal function.  

The Fourier Transform is one of the most utilized mathematical transforms in science and engineering.  By definition, a Fourier Transform will take a given function and represent it by a series of sinusoidal functions of varying frequencies and amplitudes.  Analytically, the spectral function $F(\omega)$ is represented as \cite{PDEbook,ODEbook} 
\begin{eqnarray}
\label{eq:eqFourierAnalytic}
F({\omega})&=&{\int_{-{\infty}}^{\infty}}{f(t)}{\cdot}{e^{-2{\pi}{\cdot}i{\cdot}{t}{\cdot}{\omega}}}{dt},
\end{eqnarray}
where \emph{i} is the imaginary term ($i={\sqrt{-1}}$), \emph{f(t)} is any temporal function of \emph{t} to be transformed, and $\omega$ (rad/s) represents the frequency of each sinusoidal function.  The inverse of this function is 
\begin{eqnarray}
\label{eq:eqInvFourierAnalytic}
{f(t)}&=&{\int_{-{\infty}}^{\infty}}{F({\omega})}{\cdot}{e^{2{\pi}{\cdot}i{\cdot}{t}{\cdot}{\omega}}}{d{\omega}}. 
\end{eqnarray}
Conceptually, the spectral function $F(\omega)$ represents the amplitudes of a series of sinusoidal functions of discrete frequencies $\omega_k$ (rad/s)
\begin{eqnarray}
\label{eq:eqFourierSeries}
{f(t)}&=&{\Sigma_{k=0}^{\infty}}{F({\omega_k})}{\cdot}{sin({{\omega_k}{\cdot}{t}})}. 
\end{eqnarray}


Often in practical application, one does not have an exact analytical function, but a series of discrete data points at discrete times $t_n$.  If it is necessary to convert this discrete data into the spectral domain, the traditional approach has been to use the \emph{Discrete Fourier Transform} (DFT) algorithm, often known as \emph{Fast Fourier Transform} (FFT).  The DFT algorithm is, by definition \cite{CompPhys,OptsMatlab}
\begin{eqnarray}
\label{eq:eqFFT}
{F(\omega_k)}&=&{\Sigma_{n=0}^{N-1}}{f(t_n)}{\cdot}{e^{-2{\pi}{\cdot}{i}{\cdot}{\omega_k}{\cdot}{n}/N}}.
\end{eqnarray}
where $F(\omega_k)$ is a discrete spectral data point, and $f(t_n)$ is a discrete data point in the temporal domain.  With DFT, the spectral resolution is proportional to the temporal resolution, and it is often the case that the limited temporal data will not be sufficient to obtain the spectral resolution desired.  

\begin{figure}
\centering
\includegraphics[width=\textwidth]{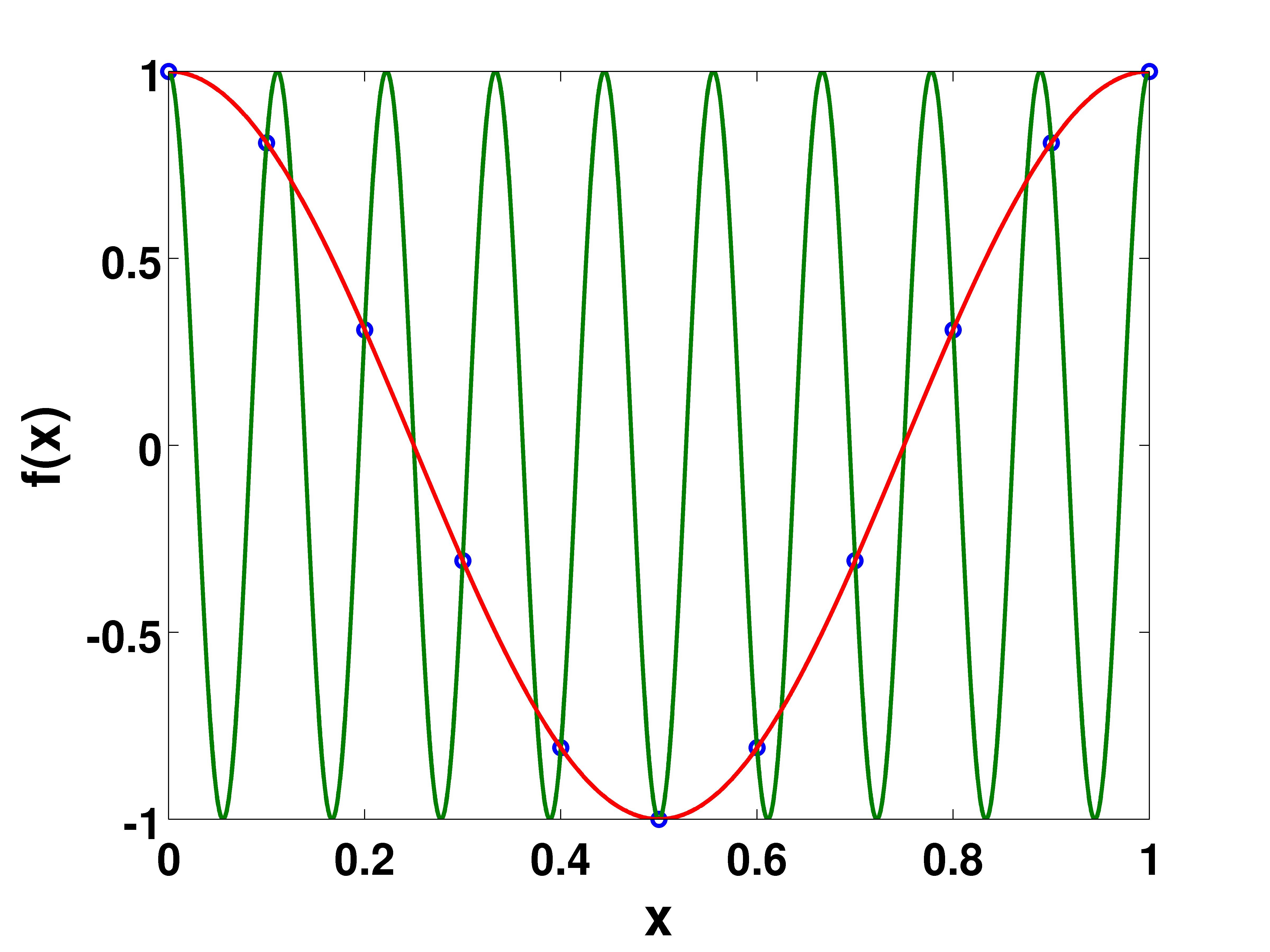}\\
\caption{Equal values for \emph{f=1} and \emph{f=9} for $cos({2}{\pi}{\cdot}f{\cdot}{x})$.}
\label{fig:figNyquist_1_9}
\end{figure}

If one wants to obtain frequency information, there is a certain minimum temporal resolution necessary to properly distinguish the frequencies; this is known as the Nyquist rate \cite{Nyquist01,Nyquist02,Nyquist03,Nyquist04,Nyquist05}.  
\begin{eqnarray}
{\delta}f&=&\frac{0.5}{{\delta}t}
\end{eqnarray}
As demonstrated in Table \ref{tb:tbNyquist_1_9} and Figure \ref{fig:figNyquist_1_9}, two different cosine functions with frequencies of \emph{1} and \emph{9} have exactly the same results when resolved at a temporal resolution ${\delta}{t}=0.1$.  

\begin{table}[h]
\begin{center}
\begin{tabular}{ | c | c | c |}
  \hline
{x} & {f=1} & {f=9}\\
\hline
\hline
0.0 & 1.0000 & 1.0000\\
\hline
0.1 & 0.8090 & 0.8090\\
\hline
0.2 & 0.3090 & 0.3090\\
\hline
0.3 & -0.3090 & -0.3090\\
\hline
0.4 & -0.8090 & -0.8090\\
\hline
0.5 & -1.0000 & -1.0000\\
\hline
0.6 & -0.8090 & -0.8090\\
\hline
0.7 & -0.3090 & -0.3090\\
\hline
0.8 & 0.3090 & 0.3090\\
\hline
0.9 & 0.8090 & 0.8090\\
\hline
1.0 & 1.0000 & 1.0000\\
\hline
\end{tabular}
\caption{Equal values for \emph{f=1} and \emph{f=9} for $cos({2}{\pi}{\cdot}f{\cdot}{x})$.}
\label{tb:tbNyquist_1_9}
\end{center}
\end{table}

There are many approaches to implementing Fourier transforms on data of limited resolution.  One method is to introduce a scaled coordinate system and identifying the Fourier variables as the direction cosines of propagating light have been used to spectrally characterize diffracted waves in a method known as Angular Spectrum Fourier transform (FFT-AS) \cite{HarveyFFT1979,Stamnes01,Stamnes02,Stamnes03}.  Another technique of numerical Fourier Transform is Direct Integration (FFT-DI) \cite{AO_RSd2006}, using Simpson's rule to improve the calculations accuracy.  Finally, one of the simplest approaches to taking the Fourier transform with a limited temporal resolution is to use Non-uniform Discrete Fourier Transforms (NDFT) \cite{VarFFT01, VarFFT02, VarFFT03, VarFFT04, VarFFT05, VarFFT06, VarFFT07, VarFFT08, VarFFT09}
\begin{eqnarray}
\label{eq:eqNDFT}
{F(\omega_k)}&=&{\Sigma_{n=0}^{N-1}}{f(t_n)}{\cdot}{e^{-2{\pi}{\cdot}{i}{\cdot}{p_n}{\cdot}{{\omega}_k}}}.
\end{eqnarray}
where $0<{p_n}<1$ are relative sample points over the range, and ${\omega}_k$ is the frequency of interest.

\section{{Spectral Transform Algorithm}}

This algorithm, which the author calls the \emph{Coefficient of determination Fourier Transform} (CFT), is an approach to obtain greater spectral resolution; the full spectral domain, or any frequency range or resolution desired, is determined by the user.  Greater resolution or a larger domain will inherently take longer to solve, depending on the computer resources available.  One advantage of this approach is that the spectral domain can also have varying resolutions, for enhanced resolution at points of interest without dramatically increasing the computation cost of each spectral transform.  

At each discrete point in the spectral domain, the algorithm generates two sinusoidal functions
\begin{eqnarray}
\label{eq:eqSinCos}
{\Phi_k(t)}&=&{A}{\cdot}cos(2{\pi}{\cdot}{\omega_k}{\cdot}{t}),\\ \nonumber
{\hat{\Phi}_k(t)}&=&{A}{\cdot}sin(2{\pi}{\cdot}{\omega_k}{\cdot}{t}),\nonumber
\end{eqnarray}
where ${\Phi_k(t)}$ is to represent the real spectral components, ${\hat{\Phi}_k(t)}$ is to represent the imaginary spectral components, ${\omega_k}$ is the discrete frequency of interest, \emph{t} is the independent variable of the data of interest, and \emph{A} is the amplitude of the function,
\begin{eqnarray}
\label{eq:eqAmp}
{A}&=&{max\{f(t)\}}-{min\{f(t)\}},
\end{eqnarray}
defined and the total range within the temporal data.  

\begin{verbatim}
  % MatLab code of the CFT algorithm

  % FFO is the function in the temporal domain.  Sfct is a discrete function
  % of frequencies, selected by the user, to define the spectral domain

  FFavg=mean(FF0);         FFstd=max(FF0)-min(FF0);

  for ii=2:ct
      sinfct=sin(2*pi*t*(Sfct(ii)));	 % Real cosine function
      cosfct=cos(2*pi*t*(Sfct(ii)));  % Imaginary sine function
      corrR=R2fct(cosfct,FF0);        % R^2 of real component
      corrI=R2fct(sinfct,FF0);        % R^2 of imaginary component
      SpecFct(ii)=corrR+(i*corrI);    % Saving the Spectral Function
  end
  SpecFct=FFstd*SpecFct/(sum(abs(SpecFct))); % Normalize the spectral function
  SpecFct(1)=FFavg; % Set the average of the temporal function (Sfct(1)=0)
\end{verbatim}

The next step is to take each of these functions, and find the \emph{Coefficient of Determination} (CoD) between the function and the temporal data, all with the same temporal domain and resolution \cite{R2_1,R2_2,R2_3, a15622, LinAlgebra}.  The CoD is a numerical representation of how much variance can be expected between two functions.   To find the CoD between two equal-length discrete functions $G(t_n)$ and $H(t_n)$, three coefficients are first calculated
\begin{eqnarray}
{SS_t}&=&{\Sigma_{n=1}^N}{({G(t_n)-{\bar{G}}})}{\cdot}{({H(t_n)-{\bar{H}}})},\\
{SS_1}&=&{\Sigma_{n=1}^N}({G(t_n)-{\bar{G}}})^2,\nonumber \\
{SS_2}&=&{\Sigma_{n=1}^N}({H(t_n)-{\bar{H}}})^2,\nonumber 
\end{eqnarray}
where \emph{N} is the discrete length of the two functions, and ${\bar{G}}$ and ${\bar{H}}$ represent the arithmetic mean value of functions ${G(t_n)}$ and ${H(t_n)}$.  The CoD, represented as \emph{R}, is then determined as
\begin{eqnarray}
\label{eq:eqCoD}
{R}&=&{\frac{SS_t}{\sqrt{{SS_1}{\cdot}{SS_2}}}},
\end{eqnarray}
and the closer the two functions match, the closer the value of the CoD reaches \emph{R=1}.  If there is no match at all, the CoD will be equal to \emph{R=0}, and if the two functions are perfectly opposite of each other ($G(t_n)=-H(t_n)$), the CoD goes down to \emph{R=-1}.  In practice, the CoD is often represented as the $R^2$ value
\begin{eqnarray}
\label{eq:eqR2}
{R^2}&=&{\frac{SS_t^2}{{SS_1}{\cdot}{SS_2}}}.
\end{eqnarray}
This process is repeated for every sine and cosine function generated with each frequency within the spectral domain.  The coefficients of determinations can be used to represent the spectral values, both real (cosine function) and imaginary (sine functions), for the given discrete frequency point.  These functions of \emph{R} values for the real and imaginary components are then normalized to the maximum real and imaginary values, and multiplied by the amplitude \emph{A} determined in equation \ref{eq:eqAmp}.  The final outcome is a phase-resolved spectral transformation of the input function, but with a spectral domain as large or resolved as desired.  

\begin{verbatim}
  % Correlation Coefficient Function
  function [R2]=R2fct(X,Y)

  ctx=length(X);       foo=zeros(ctx,3);
  for ii=1:ctx
      foo(ii,1)=(X(ii)-(mean(X)))*(Y(ii)-(mean(Y)));
      foo(ii,2)=(X(ii)-(mean(X)))^2;
      foo(ii,3)=(Y(ii)-(mean(Y)))^2;
  end
  foo=sum(foo);
    
  foo2=(sqrt((foo(2))*(foo(3))));
  if foo2==0
      corr=0;
  else
      corr=foo(1)/foo2; % Closer to 1 is best
  end
  R2=corr^2;
  
  end
\end{verbatim}

Finally, this spectral transformation can easily be converted back to the temporal domain.  By definition, the temporal domain is merely the sum of the series of sinusoidal waves (equation \ref{eq:eqFourierSeries}), and thus the inverse CFT transform can simply be defined as 
\begin{eqnarray}
\label{eq:eqICFT}
{f(t)}&=&{{\Sigma}_{k=1}^N}{\{{real(F(\omega_k))}{\cdot}{cos({\omega_k}{\cdot}{t})}\}}+{\{{imag(F(\omega_k))}{\cdot}{sin({\omega_k}{\cdot}{t})}\}}.
\end{eqnarray}

\begin{verbatim}
  % MatLab code of the inverse CFT algorithm
  
  % Ftest = temporal output function, of discrete length ctT
  % t = time domain of discrete length ctT
  % SpecFct = spectral function of discrete length ct to be transformed 
  %           back to the temporal domain
    
  Ftest=zeros(ctT,1);
  for ii=1:ct
      Ftest=((real(SpecFct(ii)))*cos(2*pi*t*Sfct(ii)))+Ftest;
      Ftest=((imag(SpecFct(ii)))*sin(2*pi*t*Sfct(ii)))+Ftest;
  end
\end{verbatim}

\section{Initial Demonstration of the Spectral Transform Algorithm}

To demonstrate the capability of this algorithm, six functions are generated based on the two similar functions demonstrated in Table \ref{tb:tbNyquist_1_9} and Figure \ref{fig:figNyquist_1_9}; the two functions are used with a temporal range of 0 to 1, with the same temporal resolution of ${\delta}t=0.1$, and frequencies of both \emph{f=1} and \emph{f=9}.  The cosine functions are modified to have a phase shift of ${\pm}2{\pi}/3$.  

\begin{figure}
\centering
\includegraphics[width=\textwidth]{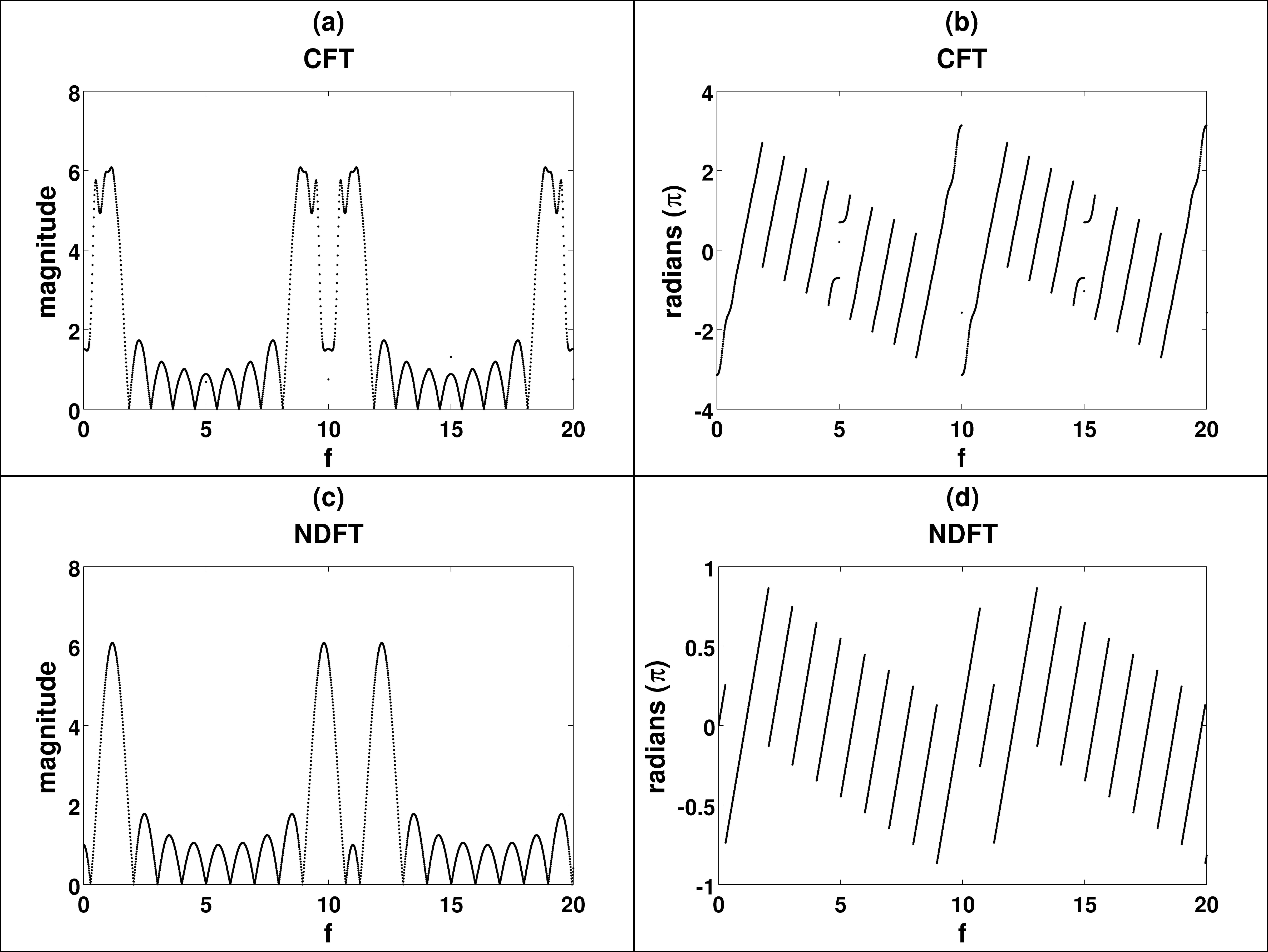}\\
\caption{Spectral results of the function $cos(2{\pi}{\cdot}x)$ with a $\delta$s=0.1, with both the proposed CFT in magnitude (a) and phase (b), as well as NDFT in magnitude (c) and phase (d).}
\label{fig:figCFFT_Sabs}
\end{figure}

The spectral transform was taken of all six of these functions with both the CFT algorithm, as well as the NDFT algorithm defined in equation \ref{eq:eqNDFT}.  The spectral magnitude and phase from both methods are plotted in Figure \ref{fig:figCFFT_Sabs}.  By using equation \ref{eq:eqICFT} to get back to the temporal domain, all six functions matched ($R^2$>0.996) with the spectral magnitude and phase from the CFT; there is no coherent match for the NDFT. This is realized by finding the coefficient of determination between the recovered temporal data and the original temporal data; the $R^2$ results are tabulated in Table \ref{tb:tbSampleFct}.  While the NDFT may give a clear picture of the spectral domain of the function, it is impossible to recover the function back to the original temporal domain without excessively computationally intensive matrix analysis.  The strength of CFT transform lies in its inverse operator defined in equation \ref{eq:eqICFT}, with which the true temporal function can be obtained back from the spectral domain obtained with highly resolved CFT.  

\begin{table}[h]
\begin{center}
\begin{tabular}{ | c | c | c | c |}
  \hline
{f (Hz)} & {Phase} & {$R^2$ CFT} & {$R^2$ NDFT}\\
\hline
\hline
1 & 0 & 0.9999 & 0.1197\\
\hline
1 & 2${\cdot}\pi$/3 & 0.9967 & 0.0091\\
\hline
1 & -2${\cdot}\pi$/3 & 0.9964 & 0.0163\\
\hline
9 & 0 & 0.9999 & 0.1197\\
\hline
9 & 2${\cdot}\pi$/3 & 0.9964 & 0.0163\\
\hline
9 & -2${\cdot}\pi$/3 & 0.9967 & 0.0091\\
\hline
\end{tabular}
\caption{Correlation between the original temporal function and the temporal function retrieved (equation \ref{eq:eqICFT}) from the spectral plot obtained with both CFT and NDFT (Figures \ref{fig:figCFFT_Sabs}). }
\label{tb:tbSampleFct}
\end{center}
\end{table}

\begin{figure}
\centering
\includegraphics[width=\textwidth]{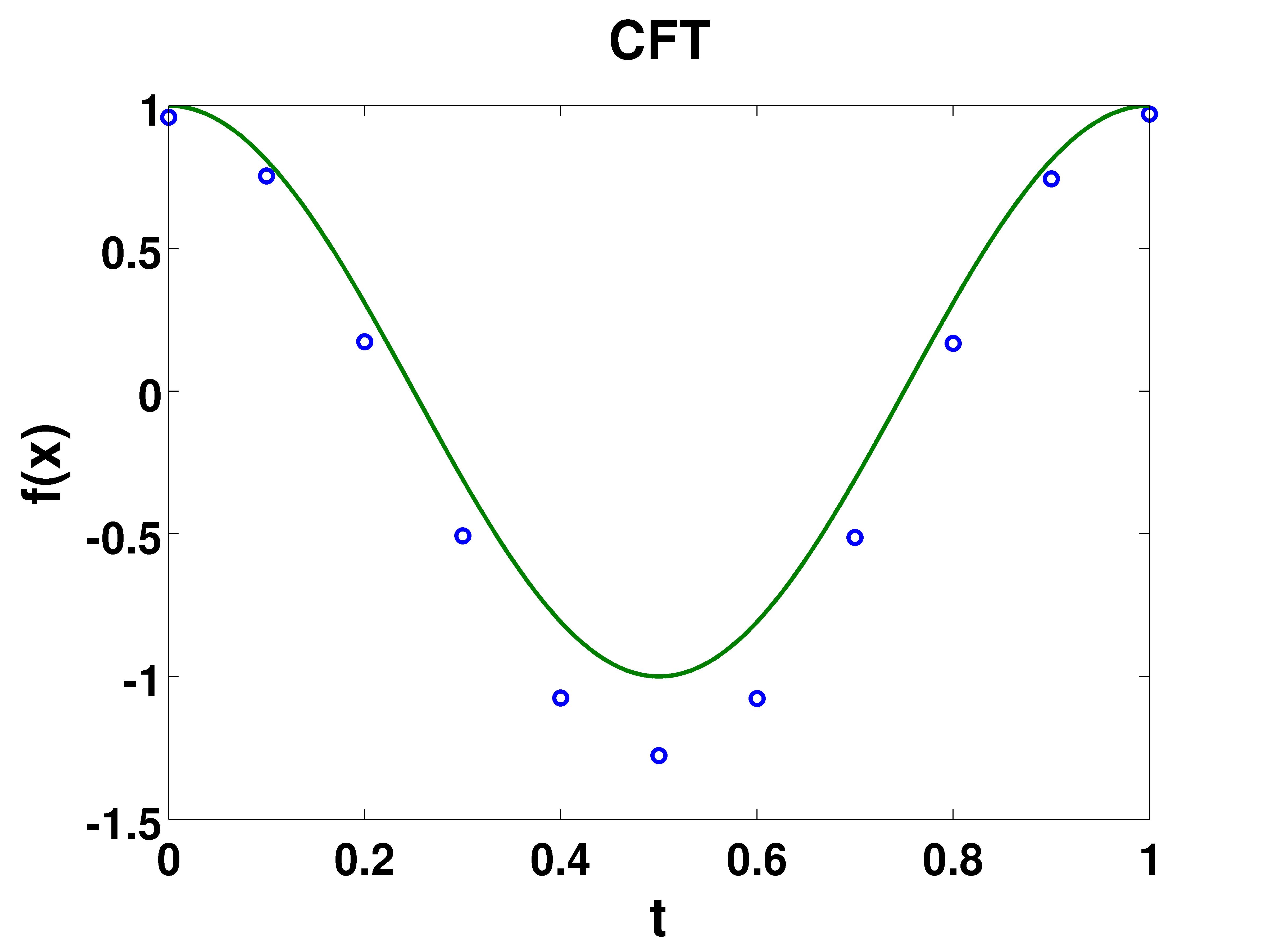}\\
\caption{Demonstration of the function $cos(2{\pi}{\cdot}{t})$, both the original function (solid green lines), and the output (blue circles) obtained from equation \ref{eq:eqICFT} and the CFT spectral results, obtained with a limited initial temporal resolution of ${\delta}{t}=0.1$.  }
\label{fig:figt_CFFT}
\end{figure}

One important point, while the CFT transform is a spectral representation of the transformed temporal function, the spectral function of this method is \emph{not} a true Fourier transform defined analytically in Equation \ref{eq:eqFourierAnalytic}.  Both the two functions are similar (see Figure \ref{fig:figCFFT_Sabs}), but not exact; this function, by definition, is the magnitude of the correlation coefficient $R^2$ of the temporal function with the associated sine and cosine wave.  For this reason, the CFT versions of the spectral representation, while similar, are \emph{not} identical to the NDFT versions, and thus are not true Fourier transforms.  The power of these CFT spectral plots lie in the fact that they can be easily transformed back to the temporal domain.  

\section{{Performance of the Spectral Transform Algorithm}}

\begin{figure}
\centering
\begin{minipage}[b]{0.30\linewidth}
\centering
\includegraphics[width=\textwidth]{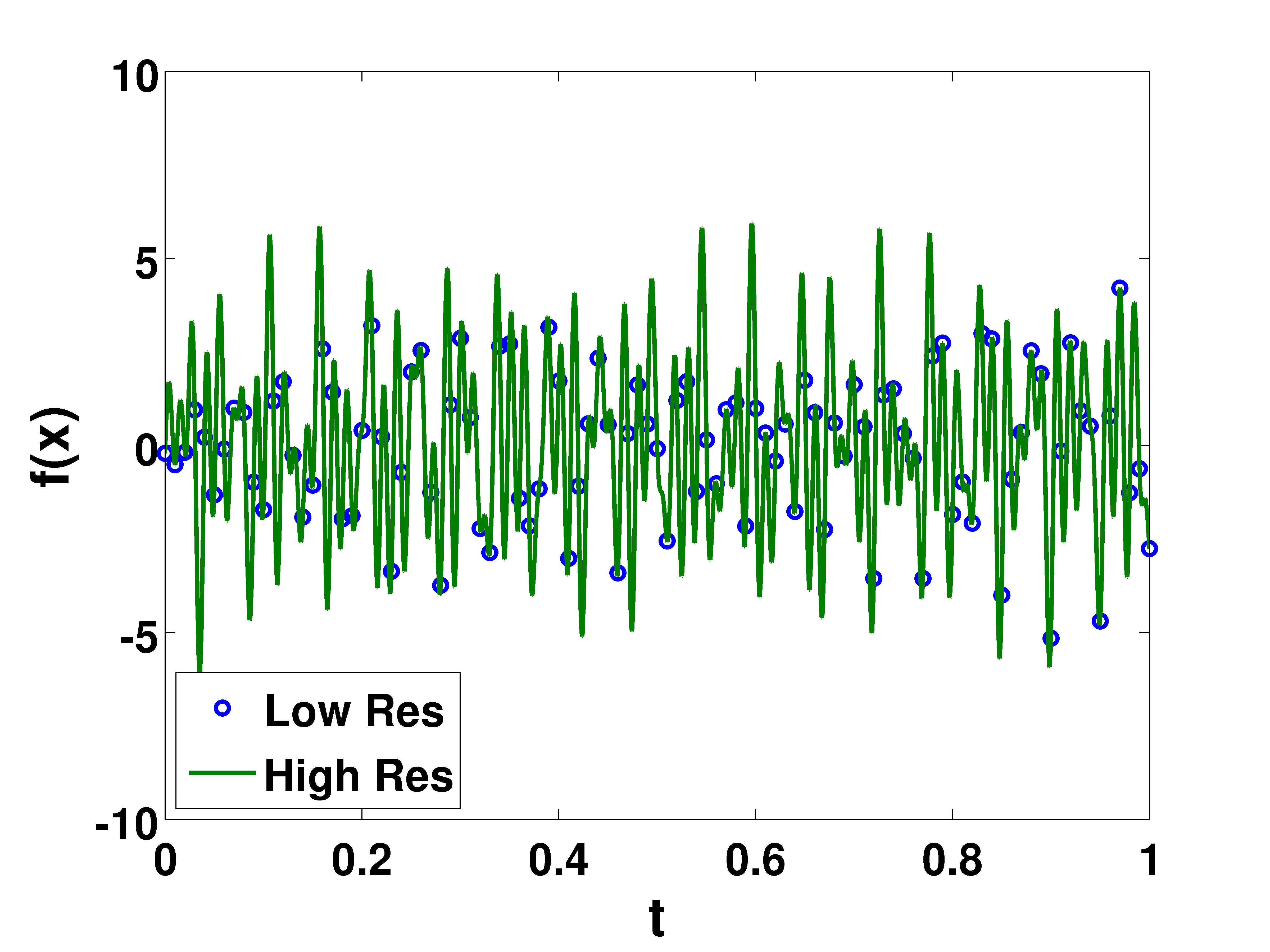}\\
\textbf{(a)}
\end{minipage}
\begin{minipage}[b]{0.30\linewidth}
\centering
\includegraphics[width=\textwidth]{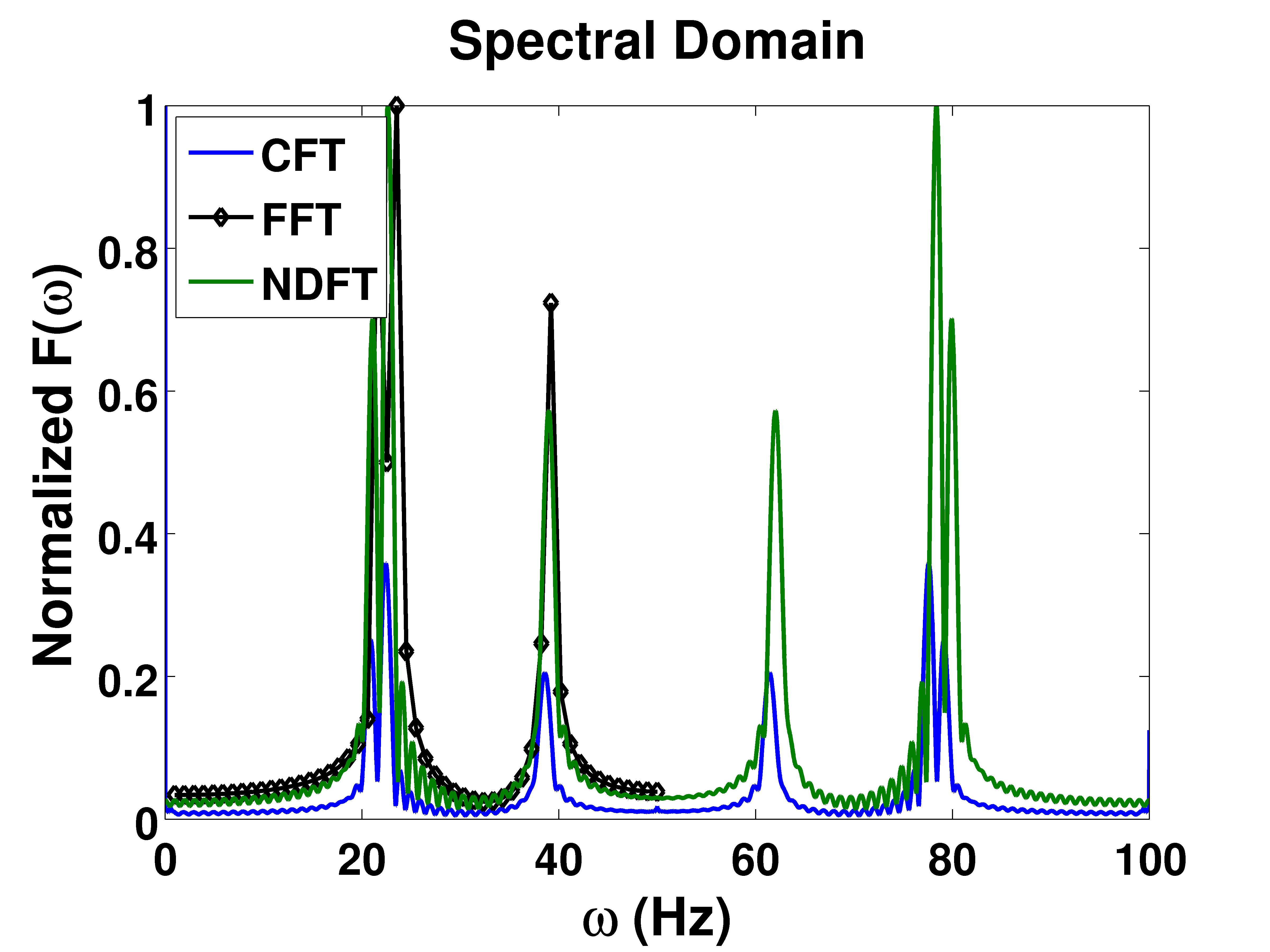}\\
\textbf{(b)}
\end{minipage}
\begin{minipage}[b]{0.30\linewidth}
\centering
\includegraphics[width=\textwidth]{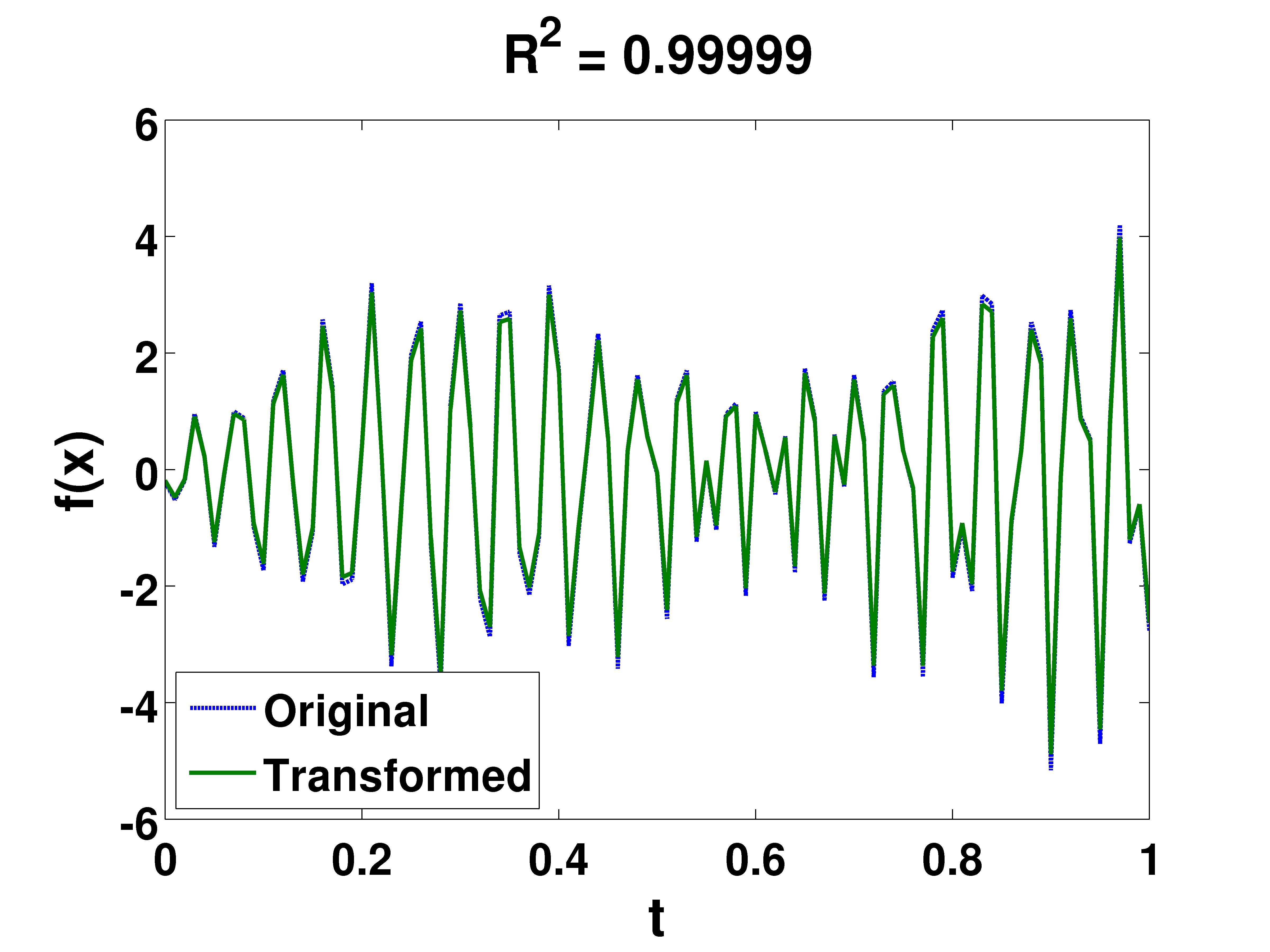}\\
\textbf{(c)}
\end{minipage}
\caption{Performance Data, for an input temporal plot of 101 data points of resolution.  (a) A comparison of the data to the true, high resolution function.  (b) The Spectral function, both the CFT and a tradition FFT analysis.  (c) The original versus the inverse transform of the spectral function.  }
\label{fig:fig_Perform_101}
\end{figure}

In the previous sections, this algorithm was demonstrated to be an effective tool to determine spectral magnitude and phase.  Next, this algorithm’s performance was compared to both traditional FFT \cite{CompPhys,OptsMatlab} of limited resolution, as well as the higher resolution NDFT method \cite{VarFFT01, VarFFT02, VarFFT03, VarFFT04, VarFFT05, VarFFT06, VarFFT07, VarFFT08, VarFFT09}.  Often in practical applications it is necessary to determine the frequencies of the peak spectral magnitudes with temporal data of limited resolution.  In this example, a temporal function is comprised of four sinusoidal functions with an amplitude of 1.25, 1.5, 1.75, and 2, at frequencies of 20.80 Hz, 38.38 Hz, 61.38 Hz, and 77.55 Hz, at a phase shift of 0$^{\circ}$, 120$^{\circ}$, 240$^{\circ}$, and 0$^{\circ}$ respectively.  Mathematically the function can be described as
\begin{eqnarray}
\label{eq:eqDummyFct}
f(t)&=&{{\frac{5}{4}}{\cdot}{cos{({2{\pi}}{\cdot}{20.80}{\cdot}t)}}}+{{\frac{3}{2}}{\cdot}{cos{({2{\pi}}{\cdot}{38.38}{\cdot}t+{{\frac{2}{3}}{\pi}})}}}+... \\ \nonumber
&&{{\frac{7}{4}}{\cdot}{cos{({2{\pi}}{\cdot}{61.38}{\cdot}t+{{\frac{4}{3}}{\pi}})}}}+{{2}{\cdot}{cos{({2{\pi}}{\cdot}{77.55}{\cdot}t)}}}
\end{eqnarray}
The function is performed over a temporal duration of 1 unit of time.  The function is plotted in Figure \ref{fig:fig_Perform_101}-a, where the green lines represent the highly resolved function, and the blue circles represent the limited temporal data of 50 Hz one might practically receive if collecting experimental data.

A traditional FFT of this limited-resolution temporal data was collected, along with a CFT with a spectral domain of 0.01 Hz ranging from 0.01 Hz to 100 Hz, as well as a NDFT transform with the same refined spectral domain; the spectral magnitudes are all plotted in Figure \ref{fig:fig_Perform_101}-b.  The peak spectral magnitudes were numerically found in the spectral range of 10-30 Hz, 30-50 Hz, 50-70 Hz, and 70-90 Hz, and tabulated in Table \ref{tb:tb_Perform_101}.  For all four frequencies, the CFT improved upon traditional FFT in terms of accurately centering on the frequency with peak spectral magnitude.  In addition, with limited resolution tradition FFT cannot even characterize frequencies higher than the resolution; the spectral plots are limited up to 50 Hz.  The NDFT method with the higher resolution allowed for spectral characterization above 50 Hz, and improved upon the accuracy of the spectral peaks when compared to the traditional FFT, but still with greater error than the CFT.  In addition, the CFT results can be accurately transformed back to the temporal domain with equation \ref {eq:eqICFT}, and in this case the transform of the spectral results matched remarkably (R$^2$=0.99999) with the original temporal function, as demonstrated in Figure \ref{fig:fig_Perform_101}-c.

\begin{table}
\begin{center}
\begin{tabular}{ | c || c | c | c |}
  \hline
  {True Frequency (Hz)} & {CFT Frequency (Hz)} & {FFT Frequency (Hz)} & {NDFT Frequency (Hz)}\\
\hline
\hline
20.8 & 22.42 & 23.5294 & 22.63\\
\hline
38.38 & 38.58 & 39.2157 & 38.98\\
\hline
61.38 & 61.42 & 50 & 62.02\\
\hline
77.55 & 77.58 & 50 & 78.37\\
\hline
\end{tabular}
\caption{Performance Results of Peak Frequency, comparing CFT versus FFT, with a temporal resolution of 101.}
\label{tb:tb_Perform_101}
\end{center}
\end{table}

\section{{Parametric Study of the Spectral Transform Algorithm}}

\begin{figure}
\centering
\includegraphics[width=\textwidth]{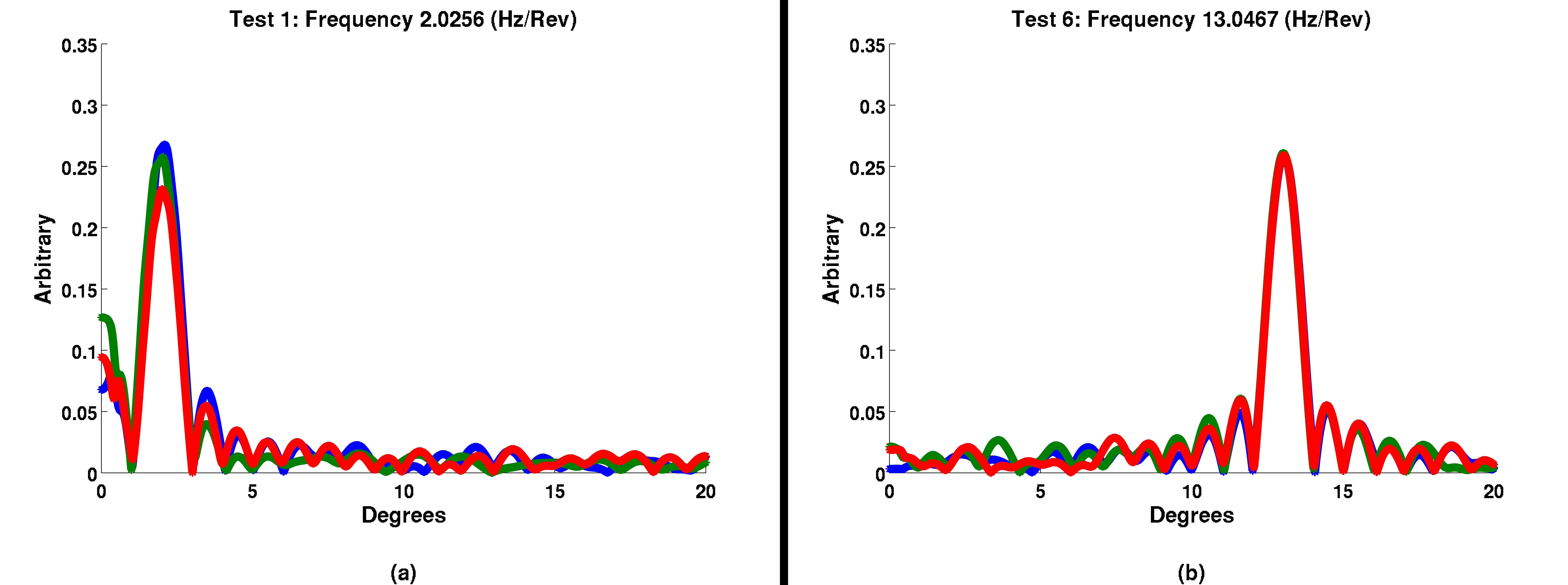}\\
\caption{Spectral results of the randomly generated functions, for frequencies of (a) 2.0256 (Hz/Rev) and (b) 13.0467 (Hz/Rev), but for different phases, magnitudes, and random noises.  }
\label{fig:figSpectralResults}
\end{figure}

A parametric study of this transform was conducted, to demonstrate that it can be used for high resolution measurements of the spectral frequency with a limited temporal resolution.  To demonstrate this, 15 random frequencies were selected, ranging from 2 to 17 cycles over the duration of the measured window.  Both the independent and dependent temporal variables are arbitrary values to demonstrate the transform function; the independent scale ranges from 0 to 1 and has 180 data points.  The arbitrary dependent data had random averages between -1000 and 1000, with an amplitude of 200 and random noise to represent the typical randomness found in typical test data.  Each of these 15 random frequencies was phase shifted by three random phases.  All forty-five arbitrary functions were transformed into the spectral domain with this transform, with a frequency domain ranging from 0 to 20 cycles per unit time duration, and a frequency resolution of 1 mHz; two examples of these spectral results are presented in Figure \ref{fig:figSpectralResults}.  As a further test of the robustness of the transform, the spectral data was then converted back to the temporal domain, and the new temporal function was compared to the original function with the coefficient of determination method to ascertain errors from the transform.  

\begin{figure}
\centering
\includegraphics[width=\textwidth]{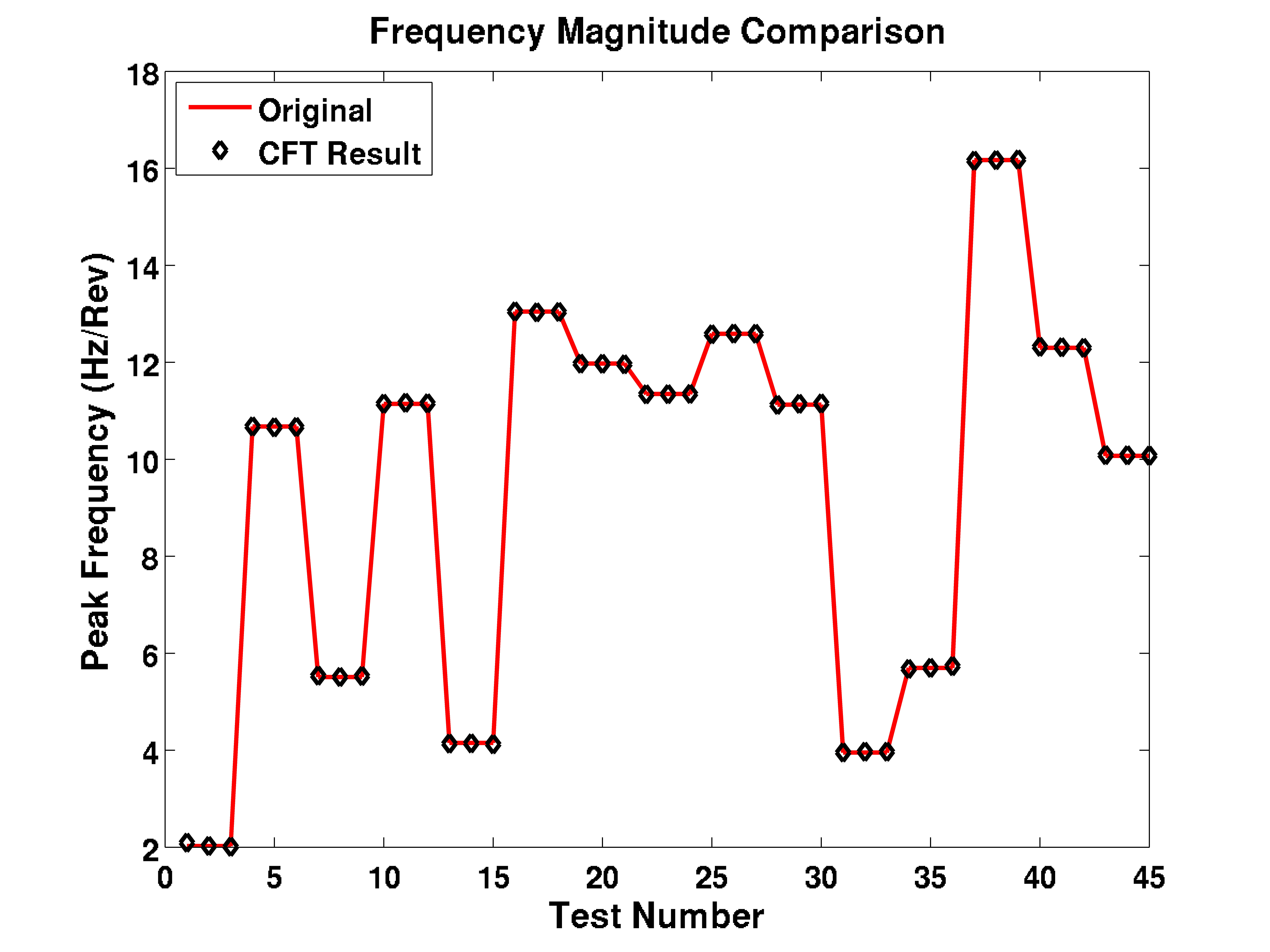}\\
\caption{Frequency Prediction Results, $R^2=0.999991$}
\label{fig:figFreqComp}
\end{figure}

\begin{figure}
\centering
\includegraphics[width=\textwidth]{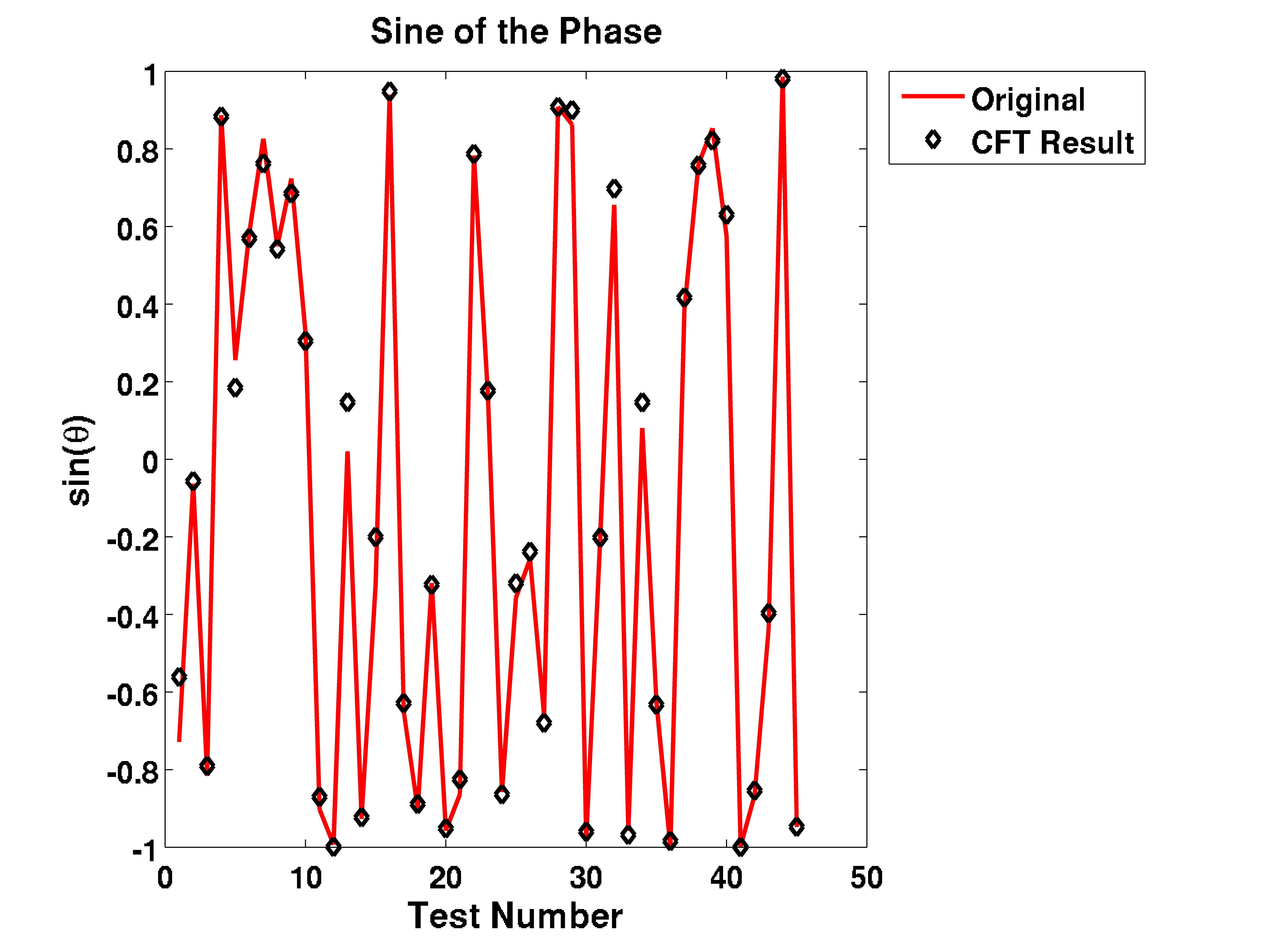}\\
\caption{Angle Prediction Results, $R^2=0.9982$}
\label{fig:figAngComp}
\end{figure}

This spectral transform was remarkably effective at finding the peak primary frequency, often with accuracy's down to tens of mHz.  The functions of the peak frequencies (Figure \ref{fig:figFreqComp}), both which were used for the initial function and the peak of the spectral transform, matches with an $R^2$ value of 0.999991; effectively identical.  The functions of the random phase angle at the peak frequencies (Figure \ref{fig:figAngComp}), both which were used for the initial function and the phase of the spectral transform at the peak frequency, matches with an $R^2$ value of 0.9982; demonstrating that this transform can be used to capture both spectral magnitude and phase with great accuracy.  

\begin{figure}
\centering
\includegraphics[width=\textwidth]{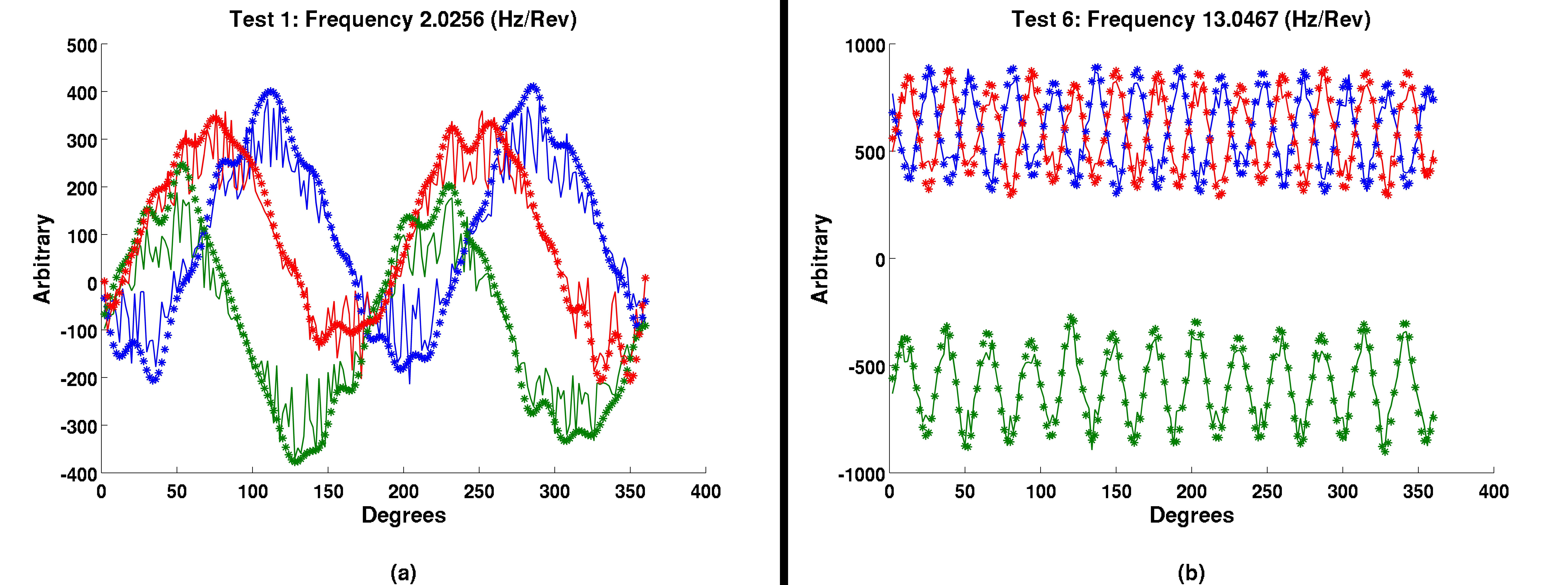}\\
\caption{Time results of the randomly generated functions, for frequencies of (a) 2.0256 (Hz/Rev) and (b) 13.0467 (Hz/Rev), but for different phases, magnitudes, and random noises.  }
\label{fig:figTimeResults}
\end{figure}

Finally, the inverse of this spectral transform was conducted for each spectral output, and the errors between the original functions and the transformed-inverse-transformed function are minimal.  As expected, not all of the fine random noise is captured; this would require a near infinite spectral domain, which would further increase computational costs, but the overarching shapes, magnitudes, and phases of the functions are consistently captured.  Taking the coefficient of determination squared of each function pair, the value of $R^2$ is never less than 0.92.  Two examples of the original function (lines) and the transformed-inverse-transformed function (stars) are represented in Figure \ref{fig:figTimeResults}.  The tabulated  results of all fifteen studies, for each of the three phase magnitude shifts, are demonstrated in Tables \ref{tb:tbResults1} - \ref{tb:tbResults3}.  

\begin{table}[h]
\begin{center}
\begin{tabular}{ | c || c | c | c | c | c |}
  \hline
{Test} & {Max Freq} & {Max Freq} & {sin(Phase)} & {sin(Phase)} & {$R^2$}\\
{ } & {Original} & {CFT Result} & {Original} & {CFT Result} & {(Temporal)} \\
\hline
\hline
1 & 2.0256 & 2.095 & -0.72837 & -0.56129 & 0.92791\\
\hline
2 & 10.6771 & 10.678 & 0.88707 & 0.88349 & 0.94843\\
\hline
3 & 5.5118 & 5.537 & 0.82648 & 0.76277 & 0.94092\\
\hline
4 & 11.1441 & 11.146 & 0.32585 & 0.30499 & 0.93117\\
\hline
5 & 4.1527 & 4.137 & 0.020562 & 0.14696 & 0.93132\\
\hline
6 & 13.0467 & 13.05 & 0.94015 & 0.94829 & 0.93359\\
\hline
7 & 11.9742 & 11.973 & -0.31888 & -0.32338 & 0.92769\\
\hline
8 & 11.3472 & 11.339 & 0.78318 & 0.78725 & 0.93398\\
\hline
9 & 12.5892 & 12.578 & -0.35645 & -0.31945 & 0.93956\\
\hline
10 & 11.128 & 11.122 & 0.90871 & 0.90812 & 0.92836\\
\hline
11 & 3.9531 & 3.954 & -0.20131 & -0.20166 & 0.93636\\
\hline
12 & 5.6981 & 5.677 & 0.080619 & 0.14678 & 0.93363\\
\hline
13 & 16.1721 & 16.158 & 0.38625 & 0.41684 & 0.94243\\
\hline
14 & 12.2989 & 12.32 & 0.57446 & 0.63023 & 0.93483\\
\hline
15 & 10.0715 & 10.085 & -0.43758 & -0.3967 & 0.92398\\
\hline
\end{tabular}
\caption{Comparison of Results, for Phase Shift Angle 1.}
\label{tb:tbResults1}
\end{center}
\end{table}
 
\begin{table}[h]
\begin{center}
\begin{tabular}{ | c || c | c | c | c | c |}
  \hline
{Test} & {Max Freq} & {Max Freq} & {sin(Phase)} & {sin(Phase)} & {$R^2$}\\
{ } & {Original} & {CFT Result} & {Original} & {CFT Result} & {(Temporal)} \\
\hline
\hline
1 & 2.0256 & 2.026 & -0.062681 & -0.056313 & 0.93311\\
\hline
2 & 10.6771 & 10.658 & 0.25575 & 0.18466 & 0.93944\\
\hline
3 & 5.5118 & 5.511 & 0.54959 & 0.54223 & 0.9434\\
\hline
4 & 11.1441 & 11.166 & -0.90216 & -0.87023 & 0.93194\\
\hline
5 & 4.1527 & 4.141 & -0.92686 & -0.92268 & 0.92885\\
\hline
6 & 13.0467 & 13.031 & -0.64986 & -0.6287 & 0.92476\\
\hline
7 & 11.9742 & 11.97 & -0.95715 & -0.95146 & 0.93294\\
\hline
8 & 11.3472 & 11.344 & 0.16368 & 0.17658 & 0.92725\\
\hline
9 & 12.5892 & 12.593 & -0.25732 & -0.23808 & 0.94473\\
\hline
10 & 11.128 & 11.145 & 0.86293 & 0.90011 & 0.93841\\
\hline
11 & 3.9531 & 3.97 & 0.65591 & 0.69795 & 0.94498\\
\hline
12 & 5.6981 & 5.7 & -0.62372 & -0.63233 & 0.93151\\
\hline
13 & 16.1721 & 16.169 & 0.76013 & 0.75789 & 0.93039\\
\hline
14 & 12.2989 & 12.31 & -0.99804 & -1 & 0.93751\\
\hline
15 & 10.0715 & 10.084 & 0.9865 & 0.98098 & 0.9411\\
\hline
\end{tabular}
\caption{Comparison of Results, for Phase Shift Angle 2.}
\label{tb:tbResults2}
\end{center}
\end{table}
 
\begin{table}[h]
\begin{center}
\begin{tabular}{ | c || c | c | c | c | c |}
  \hline
{Test} & {Max Freq} & {Max Freq} & {sin(Phase)} & {sin(Phase)} & {$R^2$}\\
{ } & {Original} & {CFT Result} & {Original} & {CFT Result} & {(Temporal)} \\
\hline
\hline
1 & 2.0256 & 2.009 & -0.81562 & -0.79053 & 0.93367\\
\hline
2 & 10.6771 & 10.666 & 0.58339 & 0.5697 & 0.92987\\
\hline
3 & 5.5118 & 5.53 & 0.72393 & 0.68535 & 0.93461\\
\hline
4 & 11.1441 & 11.154 & -0.99286 & -0.99909 & 0.92839\\
\hline
5 & 4.1527 & 4.122 & -0.3222 & -0.20013 & 0.95071\\
\hline
6 & 13.0467 & 13.04 & -0.90228 & -0.88856 & 0.94272\\
\hline
7 & 11.9742 & 11.958 & -0.86461 & -0.82531 & 0.93705\\
\hline
8 & 11.3472 & 11.347 & -0.85766 & -0.86378 & 0.93519\\
\hline
9 & 12.5892 & 12.588 & -0.66127 & -0.67959 & 0.92307\\
\hline
10 & 11.128 & 11.152 & -0.9832 & -0.96008 & 0.9276\\
\hline
11 & 3.9531 & 3.969 & -0.9574 & -0.96818 & 0.9428\\
\hline
12 & 5.6981 & 5.734 & -0.99907 & -0.9842 & 0.94156\\
\hline
13 & 16.1721 & 16.183 & 0.85337 & 0.82142 & 0.93247\\
\hline
14 & 12.2989 & 12.296 & -0.86634 & -0.85394 & 0.92086\\
\hline
15 & 10.0715 & 10.075 & -0.94753 & -0.9473 & 0.94203\\
\hline
\end{tabular}
\caption{Comparison of Results, for Phase Shift Angle 3.}
\label{tb:tbResults3}
\end{center}
\end{table}
 
\section{{Conclusion}}

This effort has demonstrated a practical, working, invertible method of numerically transforming a 1D temporal function, called \emph{Coefficient of determination Fourier Transform} (CFT), obtaining high-resolution in the spectral domain from limited resolution in the temporal domain, and retaining the ability to go back to the temporal domain from the spectral data with equation \ref{eq:eqICFT}.  The CFT algorithm, while very similar, is not a true Fourier transform as defined analytically in equation \ref{eq:eqFourierAnalytic}; it is a representation of the spectral magnitudes as related to its correlation coefficient $R^2$ value.  The CFT transform inherently is more computationally expensive than traditional DFT and NDFT methods, any desired spectral resolution and spectral domain can be used to characterize the input data; the transform can even convert the function to a spectral domain of varying resolution, so that peaks can be accurately identified without too much computational expense.  The algorithm was first tested with simple phase-shifted cosine functions, and the inverse transform of the spectral resolution matched remarkably.  Next, the spectral transform algorithm was tested at fifteen different random frequencies, all with three different random phases, all with random noises and errors, and consistently the transform was able to characterize the peak frequency and phase angle remarkably, with a higher degree of accuracy than one can expect with traditional DFT methods.  

\bibliographystyle{unsrt}
\bibliography{CFT_ref}

@article{01093941,
  title={A Fast Computational Algorithm for the Discrete Cosine Transform},
  author={Wen-Hsiung Chen and C. Harrison Smith and S. C. Fralick},
  journal={IEEE Transactions on Communications},
  volume={25},
  number={9},
  year = {September 1977},
  pages = {1004-1009}
}

@article{01455106,
  title={On the Use of Windows for Harmonic Analysis with the Discrete Fourier Transform},
  author={Fredric J. Harris},
  journal={Proceedings of the IEEE},
  volume={66},
  number={1},
  year = {January 1978},
  pages = {51-83}
}

@article{a15622,
  title={A New Least-Squares Refinement Technique Based on the Fast Fourier Transform Algorithm},
  author={Ramesh C. Agarwal},
  journal={Acta Cryst.},
  volume={34},
  year = {1978},
  pages = {791-809}
}

@article{a27715,
  title={Incorporation of fast Fourier transforms to speed restrained least-squares refinement of protein structures},
  author={Barry Finzel},
  journal={Journal of Applied Cryst.},
  volume={20},
  year = {1986},
  pages = {53-55}
}

@article{josab-17-10-1795,
  title={Spectral resolution and sampling issues in Fourier-transform spectral interferometry},
  author={Christophe Dorrer and Nadia Belabas and Jean-Pierre Likforman and Manuel Joffre},
  journal={Journal of Optics Society of America B},
  volume={17},
  number={19},
  year = {2000},
  pages = {1795-1802}
}

@article{R2_1,
  title={An R-squared measure of goodness of fit for some common nonlinear regression models},
  author={A. Colin Cameron and Frank A.G. Windmeijer},
  journal={Journal of Econometrics},
  volume={77},
  year = {1997},
  pages = {329-342}
}

@article{R2_2,
  title={R2 Measures Based on Wald and Likelihood Ratio Joint Significance Tests},
  author={Lonnie Magee},
  journal={The American Statistician},
  volume={44},
  number={3},
  year = {August 1990},
  pages = {250-253}
}

@article{R2_3,
  title={A note on a general definition of the coefficient of determination},
  author={N. J. D. Nagelkerke},
  journal={Biomelrika },
  volume={78},
  number={3},
  year = {1991},
  pages = {691-692}
}

@book{PDEbook,
    author={Richard Haberman},
    title={Applied Partial Differential Equations With Fourier Series and Boundary Value Problems, 4$^{th}$ Edition},
    year={2003},
    publisher={Prentice Hall},
    address={Upper Saddle River, New Jersey},
}

@book{ODEbook,
    author={R. Kent Nagel and Edward B. Saff and Arthur David Snider},
    title={Fundamentals of Differential Equations, 5$^{th}$ Edition},
    year={1999},
    publisher={Addison Wesley},
    address={75 Arlington Street, Suite 300 Boston, MA 02116},
}

@book{LinAlgebra,
    author={Gilbert Strang},
    title={Introduction to Linear Algebra, 3$^{rd}$ Edition},
    year={2003},
    publisher={Wellesley-Cambridge Press},
    address={7 Southgate Rd, Wellesley, MA 02482},
}

@book{CompPhys,
    author={Alejandro Garcia},
    title={Numerical Methods for Physics, Second Edition},
    year={1999},
    publisher={Addison-Wesley},
    address={75 Arlington Street, Suite 300 Boston, MA},
}

@book{OptsMatlab,
    author={Ting-Chung Poon and Taegeun Kim},
    title={Engineering Optics With Matlab},
    year={2006},
    publisher={World Scientific Publishing Co},
    address={27 Warren St, Hackensack, NJ 07601},
}

@book{MathMethodsBook,
    author={George B. Arfken and Hans J. Weber},
    title={Mathematical Methods for Physicists, Sixth Edition},
    year={2005},
    publisher={Elsevier},
    address={30 Corporate Drive, Suite 400, Burlington MA 01803},
}

@book{EngAnalysisBook,
    author={Dennis G. Zill and Michael R. Cullen},
    title={Advanced Engineering Mathematics, Second Edition},
    year={2000},
    publisher={Jones and Bartlett Publishers},
    address={Sudbury MA},
}

@book{FFTbook,
    note={ISBN 0-521-4310805},
    title={Numerical Recipies in C: The Art of Scientific Computing},
    year={1988},
    chapter={13},
    pages={584-591},
    publisher={Cambridge University Press}
}

@article{AO_RSd2006,
  title={Fast-Fourier-transform based numerical integration method for the Rayleigh–Sommerfeld diffraction formula},
  author={Fabin Shen and Anbo Wang},
  journal={Applied Optics},
  volume={45},
  number={6},
  year = {2006},
  pages = {1102-1110}
}

@article{HarveyFFT1979,
  title={Fourier treatment of near-field scalar diffraction theory},
  author={James Harvey},
  journal={American Journal of Physics},
  volume={47},
  number={11},
  year = {1979},
  pages = {974-980}
}

@article{Stamnes01,
  title={Numerical and experimental results for focusing of two-dimensional electromagnetic waves into uniaxial crystals},
  author={Daya Jiang and Jakob J. Stamnes},
  journal={Optics Communications},
  volume={174},
  year = {2000},
  pages = {321-334}
}

@article{Stamnes02,
  title={Focusing of electromagnetic waves into a uniaxial crystal},
  author={Jakob J. Stamnes and Daya Jiang},
  journal={Optics Communications},
  volume={150},
  year = {1998},
  pages = {251-262}
}

@article{Stamnes03,
  title={Numerical and asymptotic results for focusing of two-dimensional waves in uniaxial crystals},
  author={Daya Jiang and Jakob J. Stamnes},
  journal={Optics Communications},
  volume={163},
  year = {1999},
  pages = {55-71}
}

@article{VarFFT01,
  title={A Fast Algorithm for Chebyshev, Fourier, and Sine Interpolation onto an Irregular Grid},
  author={John P. Boyd},
  journal={Journal of Computational Physics},
  volume={103},
  year = {1992},
  pages = {243-257}
}

@article{VarFFT02,
  title={The type 3 nonuniform FFT and its applications},
  author={June-Yub Lee and Leslie Greengard},
  journal={Journal of Computational Physics},
  volume={206},
  year = {2005},
  pages = {1-5}
}

@phdthesis{VarFFT03,
  author       = {Alok Dutt}, 
  title        = {Fast Fourier Transforms for Nonequispaced Data},
  school       = {Yale University},
  year         = {1993},
  month        = {May},
}

@article{VarFFT04,
  title={Fast Fourier Transforms for Nonequispaced Data II},
  author={Alok Dutt and V. Rokhlin},
  journal={{SIAM} {J}ournal of {S}cientific {C}computing},
  volume={14},
  number={6},
  year = {1993},
  pages = {1368-1393}
}

@article{VarFFT05,
  title={Fast Fourier Transforms for Nonequispaced Data II},
  author={Alok Dutt and V. Rokhlin},
  journal={Applied and Computational Harmonic Analysis},
  volume={2},
  year = {1995},
  pages = {85-100}
}

@article{VarFFT06,
  title={Accelerating the Nonuniform Fast Fourier Transform},
  author={Leslie Greengard and June-Yub Lee},
  journal={{SIAM} {R}eview},
  volume={46},
  number={3},
  year = {2004},
  pages = {443-454}
}

@article{VarFFT07,
  title={Nonequispaced Hyperbolic Cross Fast Fourier Transform},
  author={Michael Dohler and Stefan Kunis and Daniel Potts},
  journal={{SIAM} {J}ournal on {N}umerical {A}nalysis},
  volume={47},
  number={6},
  year = {2010},
  pages = {4415-4428}
}

@article{VarFFT08,
  title={Nonuniform Fast Fourier Transforms Using Min-Max Interpolation},
  author={Jeffrey A. Fessler and Bradley P. Sutton},
  journal={{IEEE} {T}ransactions on {S}ignal {P}rocessing},
  volume={51},
  number={2},
  year = {2003},
  pages = {560-574}
}

@article{VarFFT09,
  title={A Nonuniform Fast Fourier Transform Based on Low Rank Approximation},
  author={Diego Ruiz-Antolin and Alex Townsend},
  journal={{SIAM} {J}ournal on {S}cientific {C}omputing},
  volume={40},
  number={1},
  year = {2018},
  pages = {529-547}
}

@article{Nyquist01,
  title={Certain Topics in Telegraph Transmission Theory},
  author={Harry Nyquist},
  journal={Proceedings of the {IEEE}},
  volume={90},
  number={2},
  year = {2002},
  pages = {280-305}
}

@article{Nyquist02,
  title={Necessary Density Conditions for Sampling and Interpolation of Certain Entire Functions},
  author={Henry J. Landau},
  journal={Acta Mathematica},
  volume={117},
  year = {1967},
  pages = {37-52}
}

@article{Nyquist03,
  title={Communication in the Presence of Noise},
  author={Claude E. Shannon},
  journal={Proceedings of the {IEEE}},
  volume={86},
  number={2},
  year = {447-457},
  pages = {1998}
}

@article{Nyquist04,
  title={The Origins of the Sampling Theorem},
  author={Hans Dieter Luke},
  journal={{IEEE} Communications Magazine},
  volume={April},
  year = {1999},
  pages = {106-108}
}

@article{Nyquist05,
  title={On the Dynamics of Automatic Gain Controllers},
  author={K. Kupfmuller},
  journal={Elektrische Nachrichtentechnik},
  volume={5},
  number={11},
  year = {2005},
  pages = {459-467}
}

\end{document}